\documentclass[%
 aip,
 amsmath,amssymb,
 reprint,%
]{revtex4-1}

\usepackage{graphicx}
\usepackage{dcolumn}
\usepackage{bm}
\usepackage{color}

\usepackage[utf8]{inputenc}
\usepackage[T1]{fontenc}
\usepackage{mathptmx}
\usepackage{etoolbox}

\newcommand{\tn}[1]{\textnormal{#1}}  

\renewcommand{\epsilon}{\varepsilon}

\makeatletter
\def\@email#1#2{%
 \endgroup
 \patchcmd{\titleblock@produce}
  {\frontmatter@RRAPformat}
  {\frontmatter@RRAPformat{\produce@RRAP{*#1\href{mailto:#2}{#2}}}\frontmatter@RRAPformat}
  {}{}
}%
\makeatother
\begin{document}

\preprint{AIP/123-QED}

\title[]{Mechanistic Framework for Multicomponent Nanoparticle Assembly: Predicting RNA-lipid and PEI-DNA nanoparticle assembly   } 
\author{Turash Haque Pial}
\affiliation{ 
Department of Materials Science and Engineering, Johns Hopkins University, Baltimore, USA
}%
\author{Tine Curk}%
\affiliation{ 
Department of Materials Science and Engineering, Johns Hopkins University, Baltimore, USA
}%
 \affiliation{Department of Physics and Astronomy, Johns Hopkins University, Baltimore, USA}
 \email{mpial1@jh.edu, tcurk@jhu.edu}

\date{\today}

\begin{abstract}
The assembly of multicomponent nanoparticles is often kinetically controlled and exhibits strong pathway dependence. Transport, solvent exchange, nucleation/growth, and collision-driven coalescence together determine not only ensemble-averaged properties but also particle-to-particle compositional heterogeneity. Here, we present a computational modeling framework for predicting nanoparticle property distributions by coupling processing conditions, early-stage self-assembly physics, and molecular chemical details with kinetic Monte Carlo (kMC) simulations. The framework couples (i) mixing conditions with solvent-exchange-mediated particle initialization and growth, and (ii) kMC simulations that resolve stochastic collision histories, electrostatics controlled coalescence, and composition at the level of individual particles. Applied to mRNA lipid nanoparticles, the model predicts size--loading correlations and provides insight into how processing-dependent assembly pathways lead to heterogeneous payload distributions. The kMC simulations further provide merging lineage histories, which explain the emergence of log-normal volume and payload distributions through multiplicative particle-growth pathways. The same framework is also applied to PEI–DNA polyelectrolyte complexation, yielding single-particle-resolved DNA–PEI stoichiometry distributions. The framework and its open-source implementation, FormLNP, provide a process-aware route to predicting and controlling single-particle property distributions across a broad range of multicomponent nanoparticle systems.
\end{abstract}

\maketitle

\section{\label{sec:level1}Introduction\protect\\ }
Mixing two or more solutions is a common route to induce nanoparticle self-assembly, enabling a wide range of technologies spanning drug delivery~\cite{mitchell_engineering_2021}, tissue engineering~\cite{kamperman_single-cell_2018}, theranostic imaging~\cite{chen_rethinking_2017}, biomolecule separations~\cite{li_selective_2020}, personal care~\cite{salvioni_emerging_2021}, and sensing and photonic applications~\cite{Sun2005CoordinationInduced}.  In these formulation processes, two or more fluid streams or solutions (containing different components) are brought together, triggering nanoparticle nucleation, growth, and ripening. As the nanoparticles form, they diffuse and collide, and these collisions can result in aggregation, complete coalescence, partial fusion, or arrested growth as the system lowers its free energy. 

Particle properties depend not only on the chemical identity of the components, but also on the coupled effects of mixing kinetics, solute and solvent diffusion, and particle–particle interactions upon collision~\cite{ahl_microfluidic_2025, shepherd_microfluidic_2021-1, zheng_preparation_2022}. 
The timescale of mixing relative to particle formation is particularly important: slow mixing creates spatially heterogeneous conditions, whereas rapid mixing tends to produce smaller and more uniform particles in some cases~\cite{pial_controlling_nodate, devos_manufacturing_2025, markwalter_design_2018}. Similarly, collision outcomes depend on the competition between diffusion and interparticle fusion kinetics; in diffusion-limited conditions, most particle encounters result in fusion, while in reaction-limited conditions, only a subset of collisions are successful~\cite{pial_controlling_nodate, roger_coalescence_2013-1, lin_buffer_2025}. This coupling between transport, growth, and composition creates a need for mechanistic models that connect processing conditions to assembly outcomes~\cite{mukherjee_understanding_2026, kim_multiphasic_2025}. Resolving single-particle pathways can therefore enable prediction of both ensemble-averaged behavior and process-dependent distributions of particle composition.

The need to predict a full population distribution is especially acute for drug carriers such as lipid nanoparticles (LNPs) and polyelectrolyte complex (PEC) nanoparticles that deliver nucleic acid payloads (mRNA, siRNA, pDNA). It is well known that their in vivo performance depends on bulk properties (e.g., average particle size) which strongly influence biodistribution~\cite{sato_relationship_2016, hassett_impact_2021, hu_size-controlled_2021}. However, single-particle features such as payload or cargo distribution heterogeneity (for example, the number of nucleic-acid copies per particle and the fraction of empty carriers) also affect drug release, efficiency, and toxicity~\cite{pial_controlling_nodate, simonsen_perspective_2024, sato_highly_2017, ndeupen_mrna-lnp_2021, bitounis_strategies_2024}. In our recent work we demonstrated that molecular dynamics (MD) and kinetic Monte Carlo (kMC) simulations can predict siRNA loading distributions across individual LNPs and provide useful design guidance for improving in vitro delivery efficiency~\cite{pial_controlling_nodate}. However, that approach was tailored to a specific LNP/siRNA system, highlighting the need for a more generalizable framework that can predict assembly outcomes across different multicomponent nanoparticle platforms.


Here, we develop a mechanistic single‑particle modeling framework for multicomponent nanoparticle assemblies that combines nucleation and growth theory with kMC to resolve pathway‑dependent formation of individual nanoparticles (Fig.~\ref{fig:methodt}). The model explicitly accounts for: (i) the chemical properties of molecules that determine the thermodynamics of self‑assembly; (ii) process conditions and kinetics (e.g., flow rate, flow‑rate ratio, solvent exchange); (iii) nucleation, growth, ripening and coalescence that occur before solutes fully mix; and (iv) payload capture during coalescence that determine per‑particle payload distributions and other single-particle properties. We demonstrate the framework in two representative systems: (1) LNPs, where solvent exchange between aqueous and organic streams drives lipid micellization and growth followed by RNA encapsulation and coalescence (Fig.~\ref{fig:epsart}A); and (2) PECs, where oppositely charged macromolecules in aqueous solution assemble via electrostatic complexation (Fig.~\ref{fig:epsart}B). We provide open-access code so this mechanistic framework can be easily adapted to model other systems of multicomponent nanoparticle assembly.

\section{\label{sec:level1}Methods\protect\\ }

\begin{figure}[!t]
    \centering
    \includegraphics[width=\columnwidth]{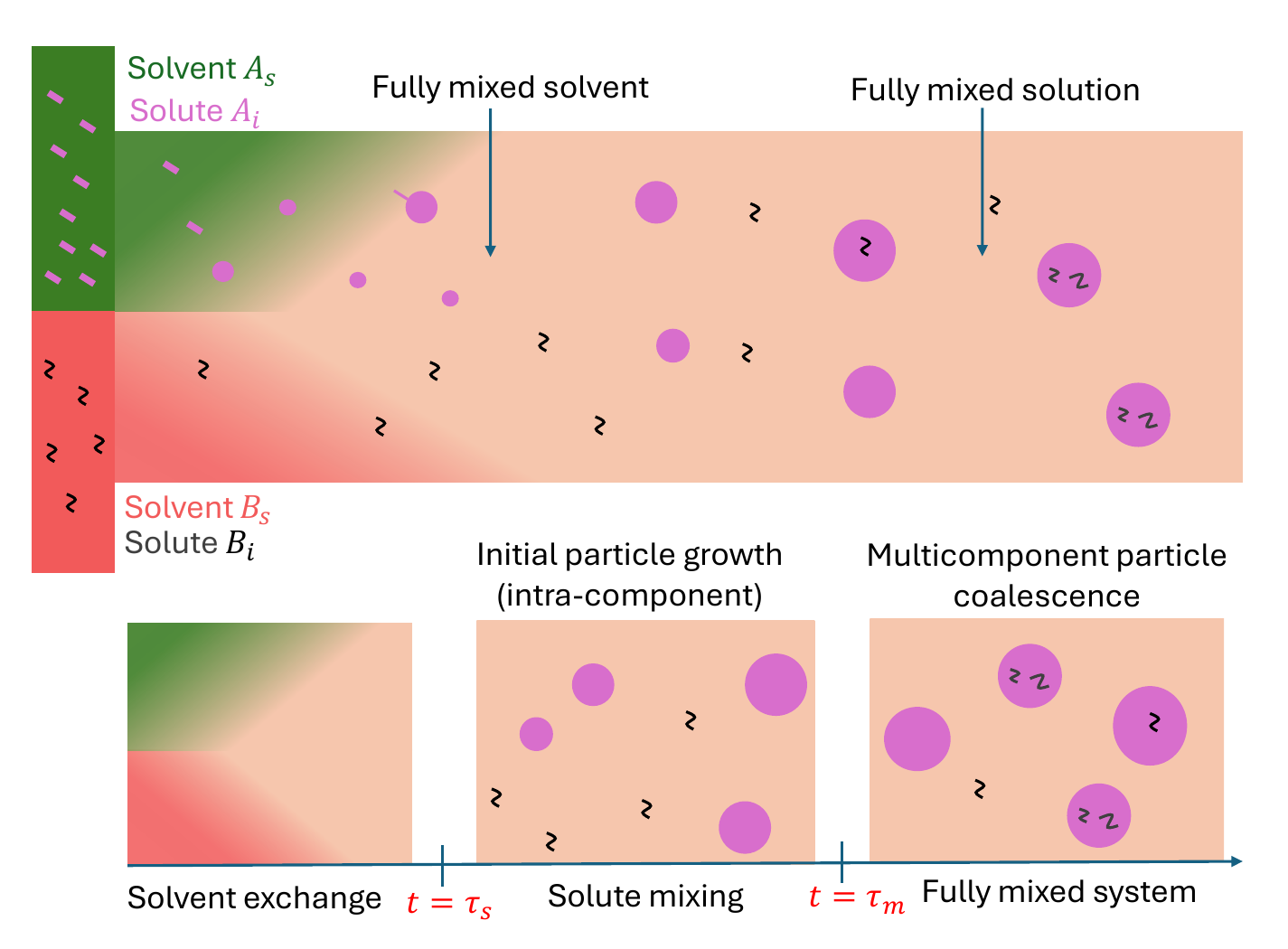}
\caption{Schematic of multicomponent nanoparticle assembly involving solvent exchange, solute diffusion, and distinct assembly stages. The lower panels illustrate the sequential stages of assembly assumed in our calculations. Solvent $A_s$ containing solutes $A_i$ and solvent $B_s$ containing solutes $B_i$ are brought into contact during mixing. Rapid solvent exchange occurs with a characteristic solvent-exchange time $\tau_s$. Solvent exchange changes the local solubility of the solutes, initiating early-stage intra-component assembly. Solutes and small nanoparticles diffuse and become fully mixed at the characteristic solute-mixing time $\tau_m$, after which collision-driven multicomponent coalescence is resolved.}
    \label{fig:methodt}
\end{figure}

We model a multicomponent assembly process in which two or more solutions (each containing different soluble components that do not self‑assemble on their own) are mixed to trigger assembly. A representative example is the co-assembly of a hydrophobic (lipophilic) solute with a hydrophilic (lipophobic) solute. Prior to mixing, each component is maintained in its respective good solvent; upon mixing, solvent exchange changes the local solubility environment and initiates self-assembly. Following this initial assembly stage, solutes originating from different solvent streams continue to diffuse and may be captured by growing particles, leading to the formation of multicomponent nanoparticles. A related mechanism occurs in electrostatic complexation, where oppositely charged polyelectrolytes are stored separately and form complexes only after mixing. In electrostatically driven complexation, nucleation and growth associated with solvent-quality changes may be absent. In this case, the assembly process can be modeled primarily as solute capture and multicomponent mixing following contact between the oppositely charged species.

In our model (Fig.~\ref{fig:methodt}), solvent exchange between solvents \(A_s\) and \(B_s\) is assumed to occur on a characteristic timescale \(\tau_s\) that is much shorter than the mixing timescale of the solutes \(A_i\) and \(B_i\). This assumption is reasonable because typical solvent molecules, such as water or ethanol, are much smaller and diffuse more rapidly than the solute species. We therefore begin the calculation after solvents \(A_s\) and \(B_s\) have effectively exchanged.

During this early stage, solute \(A_i\) is assumed, for simplicity, to form and grow into \(A_i\)-rich nanoparticles, denoted \(A_{in}\). This growth stage is continued until the \(A_{in}\) particles and the \(B_i\) solutes, which initially originate from different solvent streams, become effectively mixed at the characteristic solute-mixing time \(\tau_m\). Because growth increases the size of the \(A_{in}\) particles and thereby reduces their diffusivity, this size evolution is included when calculating \(\tau_m\).

For \(t>\tau_m\), the initially formed nanoparticles undergo binary collision and fusion events, producing larger particles. This late-stage evolution is described using coalescence kinetics with an appropriate collision kernel implemented in kMC simulations. During this stage, nanoparticles may also encapsulate solutes \(B_i\), enabling prediction of single-particle composition distributions.

\begin{figure*}[t]
\includegraphics[width=\textwidth]{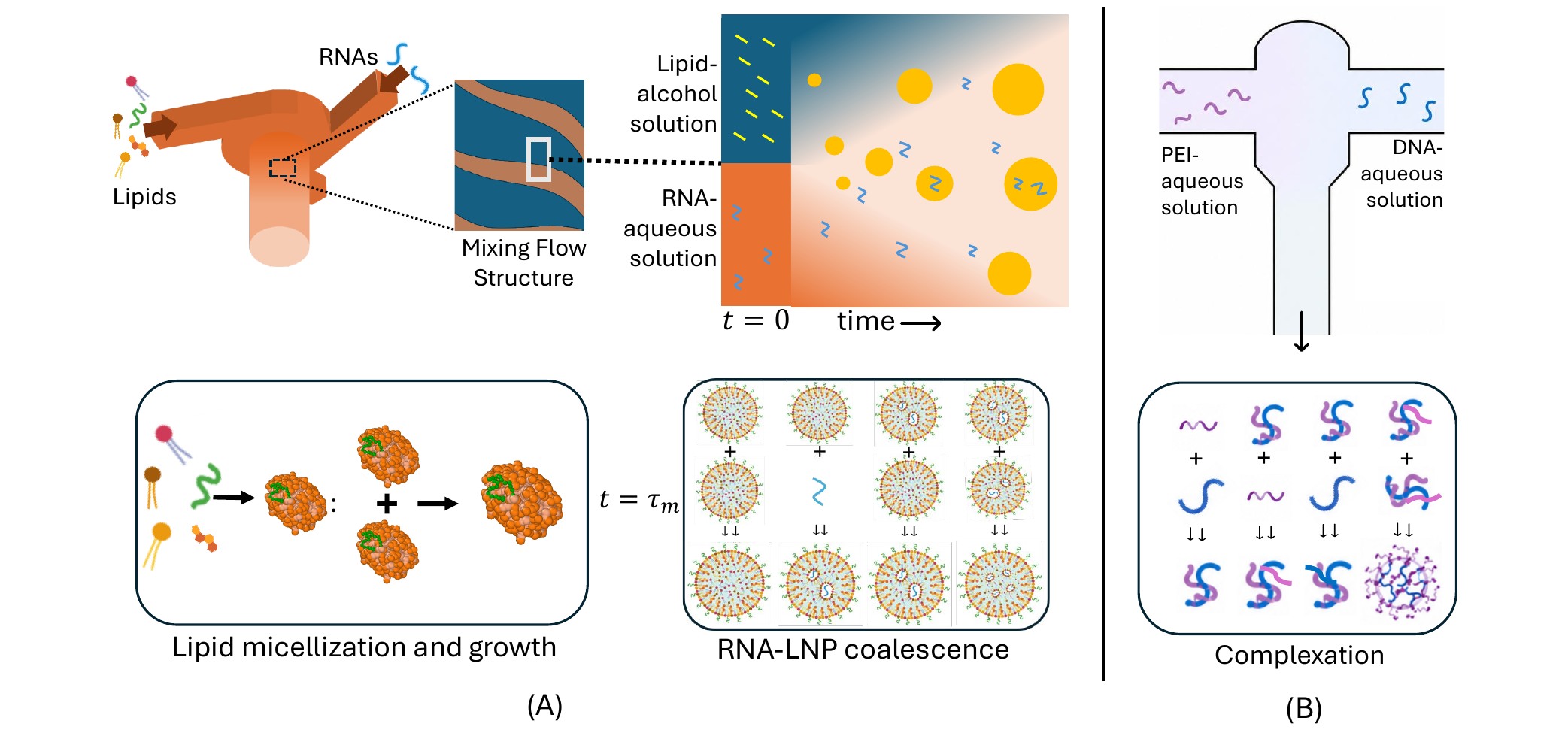}
\caption{\label{fig:epsart} Schematic of the multistage assembly pathways considered in this work. (A) For lipid nanoparticles, rapid mixing of an ethanol lipid stream with an aqueous mRNA stream creates composition gradients. These small particles subsequently undergo collision-driven coalescence with mRNA capture. (B) For polyelectrolyte complexation, oppositely charged PEI and DNA streams mix in a common aqueous solvent and form complexes directly via charge-regulated association, and cluster--cluster aggregation (no solvent-exchange effects). }
\end{figure*}

\subsubsection{\label{sec:level3}Pre-mixing Particle Growth ($t<\tau_m$)}

The early assembly stage at $t<\tau_m$ determines the particle population at the onset of stochastic coalescence. Specifically, the kMC simulations require an initial particle distribution, such as radius \(R_0\), evaluated at the characteristic solute-mixing time, \(\tau_m\). Depending on the system and desired level of resolution, this particle size and growth history can be obtained in several ways.

One option is to use molecular simulations, such as coarse-grained molecular dynamics or dissipative particle dynamics, to directly resolve solvent-exchange-mediated aggregation and early nanoparticle growth. This approach can provide molecular-scale information about early intermediates, size distributions, and pre-encapsulation of solutes~\cite{pial_controlling_nodate}. A second option is to use a nucleation-and-growth population balance model, in which solvent exchange creates supersaturation and drives particle nucleation and subsequent growth. A third option is to use a reduced merging-based growth model, which assumes initial solvent exchange-driven nucleation is very fast and initially formed particles then grow solely through particle--particle coalescence and fusion. The nucleation-and-growth population balance model and the merging-based growth model are both described in Appendix~\ref{app:growth_models}.

\subsubsection{\label{sec:level3}Determination of $\tau_m$}

In turbulent mixing, the kinetic energy of the inlet streams breaks the jets into inter‑shearing layers or turbulent eddies as the solutions enter the mixing chamber. The thickness of these inter‑shearing layers is controlled by the flow rate: higher flow rates inject more kinetic energy, producing finer breakdown and smaller eddies, while low flow rates can produce laminar flow with little or no eddying \cite{hu_kinetic_2019-1, tennekes1972first}. In our previous work we showed that the inter‑shearing layer thickness serves as a characteristic mixing length scale \cite{pial_controlling_nodate}. The time required for solute molecules to diffuse across this length scale is an approximate measure of the characteristic mixing time, the timescale over which the solute becomes effectively mixed. This characteristic mixing time and length scale applies beyond turbulent mixing. For example, in laminar microfluidic mixing, the relevant diffusion length is set by the channel or focusing geometry rather than by turbulent eddy size.

Within the characteristic mixing length scale, solutes ${A_{i}}$ can nucleate and grow into particles ${A_{in}}$ (e.g., lipid micelles), while we assume ${B_{i}}$ remain dispersed as solutes. The combined effective diffusive path relevant for ${A_{in}}$--${B_{i}}$ encounters is therefore set by the combined distance covered by the two species:
\begin{equation}
l_{\text{RMS}}(t) = \sqrt{\langle l^2_{\text{$A_{in}$}} \rangle + \langle l^2_{\text{$B_{i}$}} \rangle}
\label{eq:L_rms}
\end{equation}
where each mean-square displacement follows:
\begin{equation}
\langle l^2_i \rangle = \int_0^t 2 D_i(t') \, dt'
\label{eq:msd}
\end{equation}

Assuming ${B_{i}}$ does not self-assemble by itself (e.g., ${B_{i}}$ represents RNA or DNA payload), the ${B_{i}}$ diffusion coefficient is constant. The ${A_{in}}$ diffusivity evolves with particle size via the Stokes--Einstein relation:
\begin{equation}
D_{\text{$A_{in}$}}(t) = \frac{k_B T}{6 \pi \eta \, R(t)}
\label{eq:stokes_einstein1}
\end{equation}
where $R(t)$ is obtained from particle growth calculation and \(k_B\) is the Boltzmann constant, \(T\) is the absolute temperature, and \(\eta\) is the solvent viscosity.

For a given flow rate and geometry, we obtain the target mixing length scale $l_m$. We then integrate Eqs.~\eqref{eq:L_rms}--\eqref{eq:stokes_einstein1} forward in time, updating the combined diffusive length until $l_{\text{RMS}}(t)$ reaches the target value:
\begin{equation}
l_{\text{RMS}}(\tau_\tn{m}) = l_m
\label{eq:t_mix_def}
\end{equation}
%
%
%
This approach couples particle size evolution with diffusive mixing, enabling prediction of both the mixing timescale ${\tau_m}$ and the corresponding average particle size $R_{\text{mix}} = R({\tau_m})$.

This method requires characteristic mixing time or length scale. For turbulent mixers, these quantities can be estimated from empirical
correlations or from the Kolmogorov length scale. In general, the characteristic mixing length may be expressed as a power-law function of the total flow rate~\cite{hu_kinetic_2019-1}:
\begin{equation}
l_m = A \times  \, Q^{-B}
\label{eq:tau_mixing}
\end{equation}
where \(Q\) is the total flow rate, and \(A\) and \(B\) are empirical constants that depend on the mixer geometry and operating conditions. In microfluidic mixing chambers, however, $l_m$ may be set primarily by the chamber dimensions rather than by the flow-rate-dependent turbulent mixing correlation.


\subsubsection{\label{sec:level3}Kinetic Monte Carlo Simulation ($t>\tau_m$)}

Following complete solute mixing, particle coalescence is simulated using kinetic Monte Carlo to track individual particle histories. The initial particle configuration, such as particle radius and composition, are taken from the pre-kMC particle growth model at the characteristic mixing time $\tau_m$.

Binary collisions are treated as merging events, allowing particles to evolve in composition according to their collision histories while preserving the identity and amount of each component. During merging, both total particle volume and the amount of each component are conserved. When particles $j$ and $k$ merge, the total volume satisfies:
\begin{equation}
V_{jk}=V_j+V_k.
\end{equation}
If $f_j^{(m)}$ and $f_k^{(m)}$ are the volume fractions of component $m$ in particles $j$ and $k$, respectively, then the composition of the merged particle is determined by:
\begin{equation}
f_{jk}^{(m)}V_{jk}=f_j^{(m)}V_j+f_k^{(m)}V_k.
\end{equation}

The rate at which two particles with radii $R_j$ and $R_k$ merge in a system of volume $V_{\text{system}}$ is: \cite{roger_coalescence_2013-1}
\begin{equation}
k_{jk} = \frac{2 k_B T}{3 \eta} \frac{(R_j + R_k)^2}{R_j R_k} \exp\left(-\frac{E_b}{k_B T}\right) \frac{1}{V_{\text{system}}}
\label{eq:kij}
\end{equation}
where $\eta$ is the solvent viscosity and $E_b$ is the fusion barrier.

The simulation employs the direct Gillespie algorithm:
\begin{itemize}
  \item Compute all pairwise event rates \(k_{jk}\).
  
  \item Draw waiting time: $\Delta t = -\ln(u)/k_{\text{tot}}$, where $k_{\text{tot}} = \sum_{j<k} k_{jk}$ and  \(u\) is a uniformly distributed random number.
  
  \item Select event pair $(j,k)$ with probability $\propto k_{jk}$.
  
  \item Execute the merger event, update time and particle properties (and any affected rates).
  \item Repeat until a termination criterion is reached. A possible termination criterion is the total elapsed time.
\end{itemize}

The fusion barrier can be defined according to the physicochemical interactions relevant to a given multicomponent colloidal system. In general, this barrier may include Derjaguin–Landau–Verwey–Overbeek (DLVO) interactions, steric repulsion from surface-bound polymers or ligands, hydration/solvation repulsion, depletion interactions, elastic deformation, and other component-specific effects. 
The total energy barrier can be written as
\begin{equation}
E_b = E_{\text{steric}} + W_{\text{DLVO}} + E_{\text{depletion}} + ...
\label{eq:barrier}
\end{equation}
For the LNP--mRNA and PEI--DNA systems considered in the Results section, we include the dominant interaction contributions relevant to each system: steric repulsion from PEGylation for LNP--mRNA formulations and DLVO interactions for both LNP--mRNA and PEI--DNA systems. 
The PEG contribution arises from the entropic penalty of excluding PEG chains from the interaction area. Assuming PEG is mobile and only present on the surface of LNPs:
\begin{equation}
E_{\text{steric}}=E_{\text{PEG}} = \frac{2\pi k_B T R_F}{3 v_0 (1 - f_w)} \frac{R_i R_j}{R_i + R_j} \left( R_i \phi_{\text{PEG},i} + R_j \phi_{\text{PEG},j} \right)
\label{eq:Epeg}
\end{equation}
where $v_0$ is the lipid molecular volume, $f_w = 0.2$ is the water volume fraction in LNPs, \cite {tesei_lipid_2024} and $\phi_{\text{PEG}}$ is the PEG-lipid mole fraction of total lipid molecules. Here, $R_F = a (n_p)^{3/5}$ is the Flory radius, where $a = 0.37$~nm is the monomer size and $n_p$ is the degree of polymerization\cite{hogg_mutual_1966, li_brush_2021, marsh_lipid_2003}. 
The mole fraction of PEG-lipid relates to its volume  fraction in the lipid phase via $\phi^{(\text{PEG})}_i = \frac{f^{(\text{PEG})}_i}
{\sum_{\ell \in \text{lipids}} f^{(\ell)}_i}$, where the sum extends over  all lipid species within the LNP.

The DLVO interaction energy comprises van der Waals attraction and electrostatic repulsion:\cite{hogg_mutual_1966}
\begin{eqnarray}
W_{\text{DLVO}}(d) &=& W_{\text{vdW}} + W_{\text{elec}} \\
W_{\text{vdW}} &=& -\frac{A_H R_i R_j}{6d(R_i + R_j)} \\
W_{\text{elec}} &=& \frac{\varepsilon_r \varepsilon_0 R_i R_j (\psi_i^2 + \psi_j^2)}{4(R_i + R_j)} \nonumber \\
&& \times \left[ \frac{2\psi_i \psi_j}{\psi_i^2 + \psi_j^2} \ln \frac{1 + e^{-d/\lambda_D}}{1 - e^{-d/\lambda_D}} \right. \nonumber \\
&& \left. + \ln \left(1 - e^{-2d/\lambda_D}\right) \right]
\end{eqnarray}
where $A_H$ is the Hamaker constant, $d$ is the surface separation, $\lambda_D$ is the Debye length, and $\psi$ is the surface potential.

For solvent mixtures, the dielectric constant is estimated using a volume-weighted mixing rule:
\begin{equation}
\varepsilon_r = \varphi_{\text{$A_s$}} \varepsilon_{\text{$A_s$}} + \varphi_{\text{$B_s$}} \varepsilon_{\text{$B_s$}}
\label{eq:dielectric}
\end{equation}
where \(\varphi_{A_s}\) and \(\varphi_{B_s}\) are the volume fractions of solvents \(A_s\) and \(B_s\), respectively. 
For example, in a water-ethanol system, $\varepsilon_{\text{w}} = 78.5$ and $\varepsilon_{\text{eth}} = 24.3$.

\subsubsection{\label{sec:level4}Charge Regulation}

Electrostatic potentials, $\psi$, in the DLVO framework arise from the coupled effects of ionizable solutes, encapsulated charged species, and screening electrolytes. For example, in RNA-LNP; ionizable lipids, denoted $A_{\mathrm{ionizable}}$, and encapsulated nucleic-acid cargo, denoted $B_{\mathrm{ionizable}}$, can both contribute to the particle charge state, while dissolved ions screen electrostatic interactions. Describing these effects requires accounting for charge-regulation equilibria \cite{podgornik2018, avni2019, curk_charge_2021}. Within the Donnan picture \cite{doi_soft_2013}, we assume a spatially constant potential $\psi_0$ pervades the nanoparticle core, yielding an effective charge density comprising contributions from the solid phase and mobile ions:
\begin{align}
\rho &= (1 - f_w) \left[\alpha_{\text{$A_{ionizable}$}} \rho_{\text{$A_{ionizable}$}} f_{\text{$A_{ionizable}$}} + 
\alpha_{\text{$B_{ionizable}$}} \rho_{\text{$B_{ionizable}$}} f_{\text{$B_{ionizable}$}}\right] \nonumber \\
&\quad + 2 f_w c_{\text{salt}} N_A \sinh\left(\frac{e_0 \psi_0}{k_B T}\right)
\label{eq:charge_density}
\end{align}
Here, $f_w$ is the trapped water volume fraction, $\rho_{\text{$A_{ionizable}$}}$ and $\rho_{\text{$B_{ionizable}$}}$ represent the intrinsic charge densities of ionizable solutes, $\alpha_{\text{$A_{ionizable}$}}$ and $\alpha_{\text{$B_{ionizable}$}}$ capture pH-dependent protonation states, and $f_{\text{$A_{ionizable}$}}$, $f_{\text{$B_{ionizable}$}}$ are volume fractions in nanoparticles. The hyperbolic sine term describes the Boltzmann distribution of monovalent salt ions at molar concentration $c_{\text{salt}}$, where $N_A$ is  Avogadro's number and $e_0$ is the elementary charge.

The potential–charge relationship follows from the Debye--Hückel solution for a charged sphere:
\begin{equation}
\frac{e_0 \psi_0}{k_B T} = \frac{4\pi R^2 \rho \ell_B}{3\left(1 + R/\lambda_D\right)}
\label{eq:psi_charge_coupling}
\end{equation}
where $R$ is the particle radius, $\ell_B$ is the Bjerrum length, and $\lambda_D$ is the Debye screening length. Self-consistency between Eqs.~\eqref{eq:charge_density} and \eqref{eq:psi_charge_coupling} is enforced numerically via Newton--Raphson iteration.

Because $\psi_0$ depends nonlinearly on several coupled parameters, including $c_{\mathrm{salt}}$, $\rho$, pH, $pK_a$, and $R$, performing this self-consistent nonlinear solve at every kMC step would be computationally expensive. We therefore use a surrogate neural-network regression model trained to reproduce the self-consistent solution. The training procedure is described in our previous work\cite{pial_controlling_nodate}; once trained, the surrogate rapidly predicts the electrostatic potential over a broad parameter space. This provides a computationally efficient way to evaluate $\psi_0$ within the kMC simulation. 

\subsubsection{\label{sec:level4}Practical Modeling Workflow}

To apply the framework to a multicomponent nanoparticle system, we first specify the formulation inputs: solute and solvent identities, concentrations, diffusivities, molecular volumes, charge or ionization properties, pH, salt concentration, solvent flow rate ratio, mixing lengthscale, and any post-processing steps such as dialysis.

For solvent-exchange-driven systems, the local solvent composition is used to estimate solubility and supersaturation, which provide the inputs to the nucleation and growth model. In systems where classical nucleation theory (CNT) is appropriate, system-specific parameters such as the nucleation-rate prefactor [Eq.~\eqref{eq:app_nucleation_rate}] and growth-rate constant  [Eq.~\eqref{eq:app_growth_rate}] can be taken from the literature or calibrated against experiments. 
For amphiphilic molecules, however, micellization or aggregation-based models may provide a more appropriate description of the early assembly step than CNT. After solvent exchange, the resulting micelles or primary aggregates can then undergo diffusion-limited encounters and coalescence. For direct complexation systems, the nucleation--growth step can be omitted, and the reacting species can be initialized directly in the kMC simulation.

Next, the characteristic mixing length and mixing time are estimated from the mixer geometry or an empirical flow-rate correlation. For laminar flow without turbulent shear layers, the mixing length is set directly by the device geometry. If particles grow during mixing, their size-dependent diffusivity is updated when calculating the mixing time. The particle population at this mixing time provides the initial condition for kMC.

In the kMC stage, we define the allowed collision rules and the relevant interaction barriers, such as electrostatic, steric, depletion, or hydration contributions. For charged systems, charge regulation should be performed to correctly compute electrostatic interactions. For systems with fixed (non-ionizable) charges, the charge-regulation step can be skipped and the exact charge can be used directly.


\begin{figure}[!t]
    \centering
    \includegraphics[width=\columnwidth]{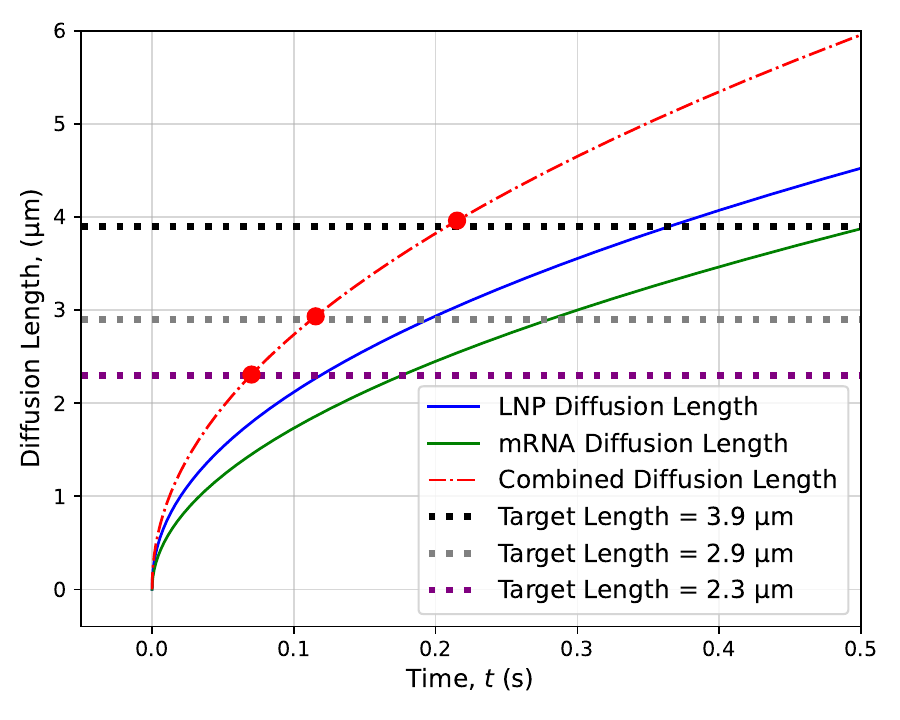}
\caption{\label{fig:interdiffusion} Evolution of RMS mixing length showing contributions from LNP and mRNA diffusion. Horizontal dotted lines mark target mixing lengths or characteristic mixing length.}
\end{figure}
\begin{figure*}[t]
\centering
\includegraphics[width=\textwidth]{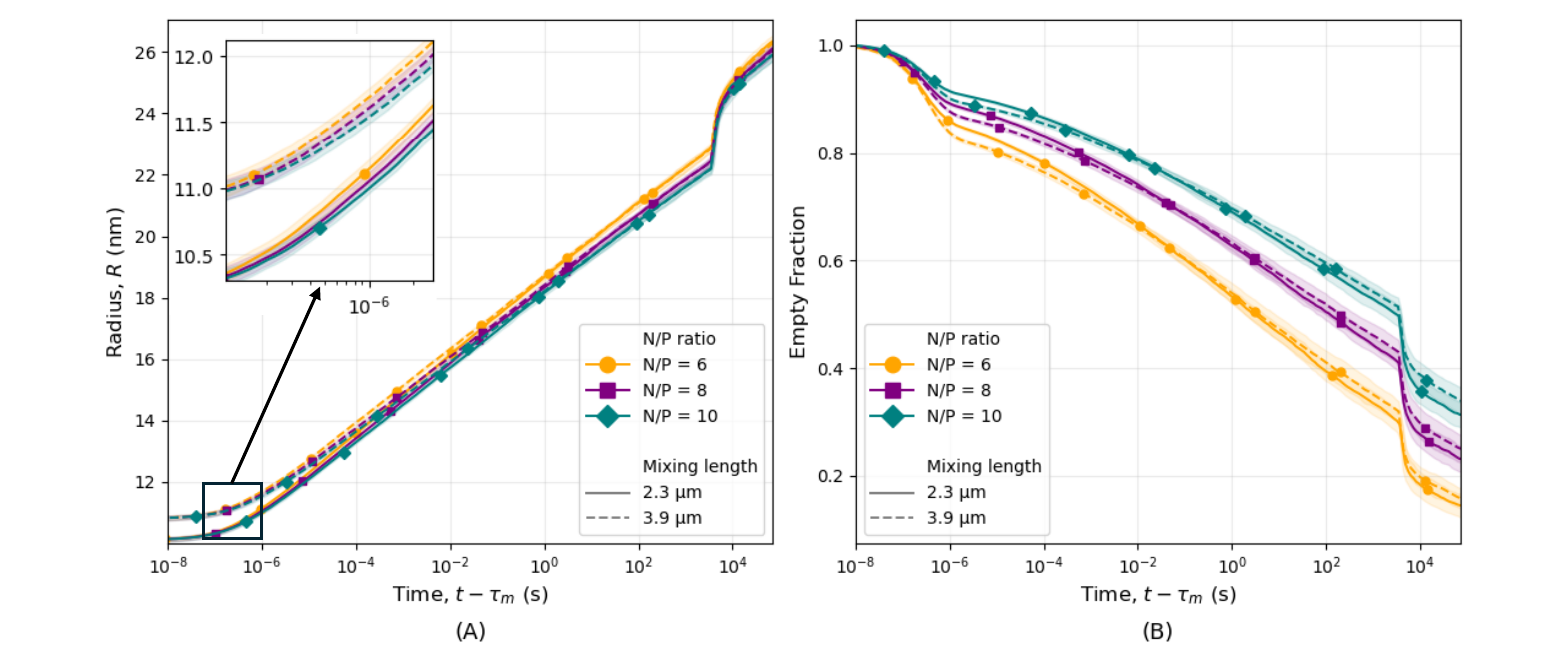}
\caption{\label{fig:kmc} (A) Average LNP radius evolution during kMC coalescence simulation for various mixing lengths and N/P (nitrogen-to-phosphorus) ratios. (B) Fraction of empty LNPs versus time, showing mRNA encapsulation kinetics. Time axis shows $t - \tau_m$, where $\tau_m$ is the mixing time. }
\end{figure*}

\subsubsection{\label{sec:level4}Limitations of the Present Model}
The present framework has several limitations that suggest clear directions for future work.
We approximate nanoparticles as spherical throughout the simulation. While this assumption simplifies diffusion and collision kernels, it does not capture anisotropic intermediates, elongated polyelectrolyte complexes, or internally phase-separated structures with blebs or irregular morphologies. Although this approximation may shift quantitative kinetics, particularly for highly aspherical particles, we expect the qualitative pathway-dependent trends and the emergence of loading heterogeneity to remain robust.

We assume that internal restructuring of nanoparticles following collision such as solute rearrangement, polyelectrolyte compaction, ion redistribution, or phase separation within the merged particle occurs rapidly compared to the collision timescale. Since the present framework focuses on compositional distributions rather than detailed internal structures, we expect that neglecting restructuring has a minimal effect on the predicted payload and composition heterogeneity. Incorporating explicit restructuring kinetics into the kMC event set would improve quantitative fidelity for predictions of particle morphology and internal organization, but is not essential for capturing composition-dependent trends.

We assume that at the characteristic mixing time $\tau_m$, the solution is fully mixed and composition gradients are negligible. In reality, although the RMS diffusive displacement reaches the target mixing length at $\tau_m$, concentration gradients will persist. These residual gradients could lead to local variations in supersaturation, nucleation rates, and particle properties. A more detailed treatment would couple the present framework to computational fluid dynamics or experimental concentration-field measurements to resolve spatial heterogeneity. For most practical formulations, however, the characteristic mixing time provides a useful transition point between the pre-mixed regime and the mixed regime.

\section{\label{sec:level1}Results\protect\\ }
\subsubsection{\label{sec:level3}Results for mRNA-LNPs}
We model an mRNA LNP formulation produced by turbulent mixing of an aqueous mRNA stream with a lipid-in-ethanol stream [Fig.~\ref{fig:epsart}A]. The lipid composition used in the simulations is DLin-MC3-DMA:DSPC:cholesterol:DMG-PEG2000 at a molar ratio of 50:10:38.5:1.5, dissolved in 100\% ethanol. We vary the total flow rate and the nitrogen-to-phosphorus (N/P) molar ratio, defined as the ratio of cationic lipids to anionic RNA. We also model a dialysis step at $t=1$ h, during which the solution pH is shifted from 4 to physiological pH. The mRNA is assumed to be 2000 nt long. Unless otherwise stated, the lipid concentration in the ethanol phase is held fixed at 10~mg/mL while changing N/P ratios. The empirical constants $A$ and $B$ in Eq.~\eqref{eq:tau_mixing} are taken from Hu et al~\cite{hu_kinetic_2019-1}. They provided an empirical relationship between \(\tau_m\), and the total flow rate, \(Q\). Their reported \(\tau_m\) values were converted to \(l_m\), using the corresponding solute diffusivity.  For the mRNA–LNP systems, we assumed a monodisperse merging-based LNP growth model, as discussed in the Appendix. This model provides the time-dependent radius of mRNA-free LNPs.

\begin{figure*}
\centering
\includegraphics[width=\textwidth]{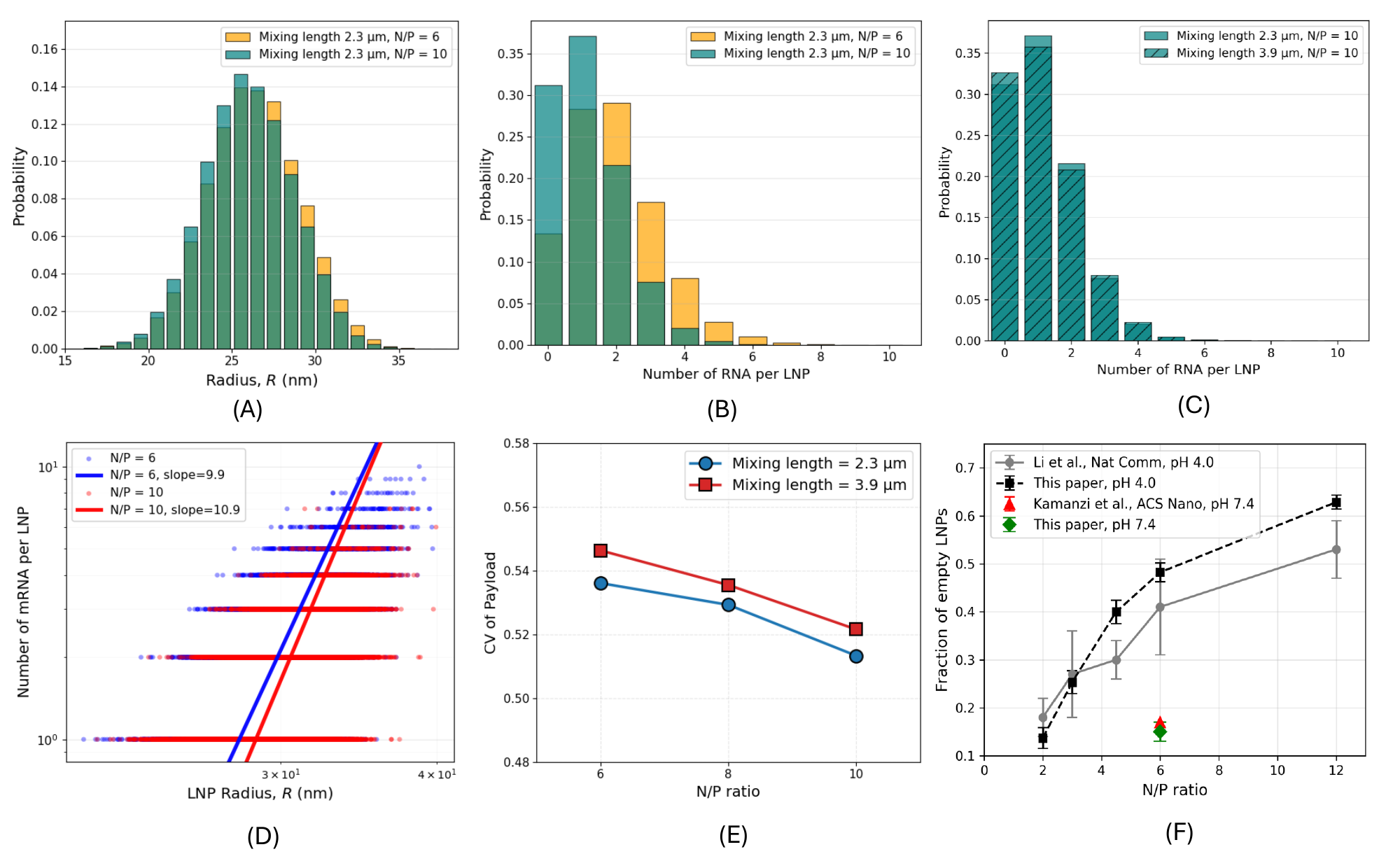}
\caption{\label{fig:final_properties} Single-particle properties from kMC simulations for LNP assembly. 
(A) Probability distributions of LNP radius at $t=18$ h. The size distributions show a modest right shift to larger radii for smaller N/P ratio. (B) Probability distributions of the number of mRNA molecules per LNP at fixed mixing length, \(l_m=2.3~\mu\mathrm{m}\), comparing N/P=6 and N/P=10 at 18 h. The bin at zero corresponds to mRNA-free LNPs. 
(C) Probability distributions of mRNA copy number per LNP at fixed N/P=10, comparing \(l_m=2.3\) and \(3.9~\mu\mathrm{m}\) at 18 h.
(D) Joint distribution of LNP radius and mRNA copy number from kMC simulations at $t=18$ h, showing a positive size--loading correlation. Larger particles tend to contain more mRNA copies, and fitted log--log slopes indicate super-volumetric scaling of payload with particle size. 
(E) Coefficient of variation (CV) of the payload distribution as a function of N/P ratio for the two mixing lengths at 18 h.
(F) Comparison of the fraction of empty LNPs as a function of N/P ratio from experiments and current kMC simulations. Gray circles show the pH 4.0 measurements reported by Li et al.~\cite{li2022payload}, while black squares show the corresponding pH 4.0 results from this work. The red triangle shows the pH 7.4 value at N/P = 6 reported by Kamanzi et al.~\cite{kamanzi2026single}, and the green diamond shows the corresponding prediction from this work at pH 7.4.
}

\end{figure*}

\textbf{Mixing timescales.}  The mixing timescale quantifies how long it takes for multiple fluid streams to diffuse and homogenize their components (solutes). As mentioned in the method section, it corresponds to the characteristic mixing length and solute diffusivity. Figure~\ref{fig:interdiffusion} illustrates the evolution of the RMS mixing length for LNPs and mRNA during the diffusion process. This figure shows representative cases with characteristic mixing lengths of $l_{\text{m}} = 3.9$~\textmu m, $l_{\text{m}} = 2.9$~\textmu m, and $l_{\text{m}} = 2.3$~\textmu m corresponds to total flow rates of $Q=20$ ml/min, $Q=30$ ml/min, and $Q=40$ ml/min respectively~\cite{hu_kinetic_2019-1}.

\textbf{Coalescence.} Figures~\ref{fig:kmc} present kMC simulation results for LNP coalescence and mRNA encapsulation over timescales spanning from microseconds to hours ($\sim$$10^5$~s). The initial particle radius for each simulation is taken from the monodisperse merging-based LNP growth model at $\tau_{\text{m}}$.

Figure~\ref{fig:kmc} (A) shows that the average LNP radius increases from 10--11~nm at $t = \tau_m$ to $\sim$26~nm after a few hours. A deflection in the growth curves is observed at 1h corresponding to dialysis. Figure~\ref{fig:kmc}(B) tracks the fraction of empty LNPs over time. Initially, all LNPs are empty (empty fraction~=~1). As coalescence proceeds, LNPs capture free mRNA through collisions, reducing the empty fraction. Complete encapsulation (empty fraction~$\to$~0) requires extended incubation times.

At \(t=\tau_m\), \(R_0\) is smaller for smaller \(l_m\) values, as shown in the inset of Fig.~\ref{fig:kmc}(A). This trend is consistent with the shorter pre-kMC growth time expected for shorter mixing length as shown in Fig.~\ref{fig:interdiffusion}. Because the N/P ratio does not affect LNP growth before the kMC stage, similar \(R_0\) values are obtained for different N/P ratios.

In contrast, the N/P ratio has a significant impact on payload distribution at later times. The fraction of empty LNPs is smaller at lower N/P ratios, indicating fewer empty particles. This behavior arises because a lower N/P ratio corresponds to a reduced lipid-to-RNA charge ratio, so that each LNP-forming lipid population is exposed to a relatively larger amount of RNA phosphate groups, which favors RNA loading.

The mixing length scale also affects the empty-particle fraction: larger \(l_m\) values lead to a higher fraction of empty particles. This trend is consistent with previous experimental observations for siRNA--LNP systems~\cite{pial_controlling_nodate}. However, in the present mRNA--LNP case, the effect is relatively small, with differences of only a few percent when the total flow rate is varied from 20 to 40 ml/min. This suggests that, for larger nucleic-acid cargoes such as mRNA, kinetic control during mixing may have a weaker influence on empty-particle formation than for smaller cargoes such as siRNA.

Figure~\ref{fig:final_properties}(A) and (B) compares the effect of N/P ratio at fixed mixing length, \(l_m=2.3~\mu\mathrm{m}\). At N/P = 6, the LNP size distribution is slightly shifted toward larger radii relative to N/P = 10, suggesting enhanced merging/coalescence at lower N/P. This is consistent with weaker electrostatic repulsion when the N/P ratio is closer to unity. At N/P = 10, the payload distribution is dominated by empty and singly loaded particles, with the highest probability occurring at one mRNA copy per LNP. Decreasing the N/P ratio to 6 reduces the empty-particle fraction and shifts the distribution toward larger mRNA copy numbers. In particular, the median payload increases from approximately one mRNA copy per LNP at N/P=10 to approximately two copies per LNP at N/P=6. 

Figure~\ref{fig:final_properties}(C) isolates the effect of mixing length at fixed N/P=10. Increasing \(l_m\) from \(2.3\) to \(3.9~\mu\mathrm{m}\) produces a modest increase in the empty-particle fraction and a slight shift of the distribution toward lower mRNA copy numbers. However, within the range examined here, the effect of mixing length is substantially weaker than that of the N/P ratio.

Figure~\ref{fig:final_properties}(D) shows a clear positive correlation between LNP size and mRNA copy number: larger particles tend to carry more mRNA, whereas smaller particles are more frequently empty or lightly loaded. This coupling is expected because larger LNPs typically undergo more coalescence events, increasing their probability of capturing mRNA. A similar size-dependent loading trend was reported by Li et al.~\cite{li_single-particle_2024} using single-particle spectroscopic chromatography measurements of siRNA--LNPs. In that study, nucleic-acid loading increased with LNP size according to an approximate power-law relationship. When plotted on a log--log scale, the payload--size relationship exhibited scaling exponents larger than expected from simple volumetric scaling, indicating super-volumetric loading behavior. Consistent with these experimental observations, our simulations also show super-volumetric scaling between mRNA copy number and LNP radius. The fitted exponents are substantially larger than 3, indicating that payload loading does not increase simply in proportion to particle volume. This agreement supports the ability of the model to capture single-particle payload heterogeneity and size-dependent nucleic-acid loading trends observed experimentally.

The coefficient of variation (CV) of the mRNA loading distribution, defined as the standard deviation divided by the mean, quantifies payload heterogeneity across the LNP population [Fig.~\ref{fig:final_properties}(E)], excluding empty LNPs. Lower CV values correspond to more uniform mRNA loading. The CV decreases as the N/P ratio increases, indicating a narrower relative payload distribution at higher N/P ratios. For all N/P ratios considered, the larger mixing length gives a slightly higher CV, consistent with a modest increase in payload heterogeneity under slower-mixing conditions.

Figure~5(F) compares the kMC-predicted fraction of empty LNPs with existing single-particle experimental measurements obtained using mRNA cargos and LNP formulations comparable to those used in this work. For the N/P-dependent comparison, we used the data from Li et al.~\cite{li2022payload}, where empty-LNP frequencies were quantified by multi-laser cylindrical illumination confocal spectroscopy (CICS). In that study, the mRNA concentration was held fixed at 20~$\mu$g/mL, and the lipid concentration was adjusted accordingly to achieve the desired N/P ratios; which we also followed in our modeling. Their measurements at pH~4.0 show an increasing fraction of empty LNPs with increasing N/P ratio, and our results reproduce the same trend across the tested N/P ratios. We also compared our physiological pH prediction with the recent single-particle study by Kamanzi et al.~\cite{kamanzi2026single}, which examined how different lipid formulations affect mRNA loading. In that work, LNP loading was quantified by combining alternating laser excitation (ALEX) with convex lens-induced confinement (CLiC). For the formulation most comparable to ours (Onpattro analog with total flow rate of 10 ml/min), Kamanzi et al. reported an empty-LNP fraction of approximately 0.17 at N/P~=~6 after 24~h. Under the corresponding simulation condition, pH~7.4, N/P~=~6, and 24~h, our model predicts an empty fraction of approximately $0.15$, in close agreement with their measurement.

\textbf{Lineage-tree analysis of merging pathways.}
Because kMC resolves individual collision histories rather than only ensemble-averaged properties, it can connect observed distribution shapes to the underlying sequence of particle--particle merging events through lineage-tree reconstruction. Motivated by experimental findings that LNP size and RNA payload distributions often exhibit log-normal distribution~\cite{li_single-particle_2024}, we use these lineage trees to visualize merging pathways and identify the growth mechanisms that give rise to the final distributions.

\begin{figure*}[t]
\centering
\includegraphics[width=\textwidth]{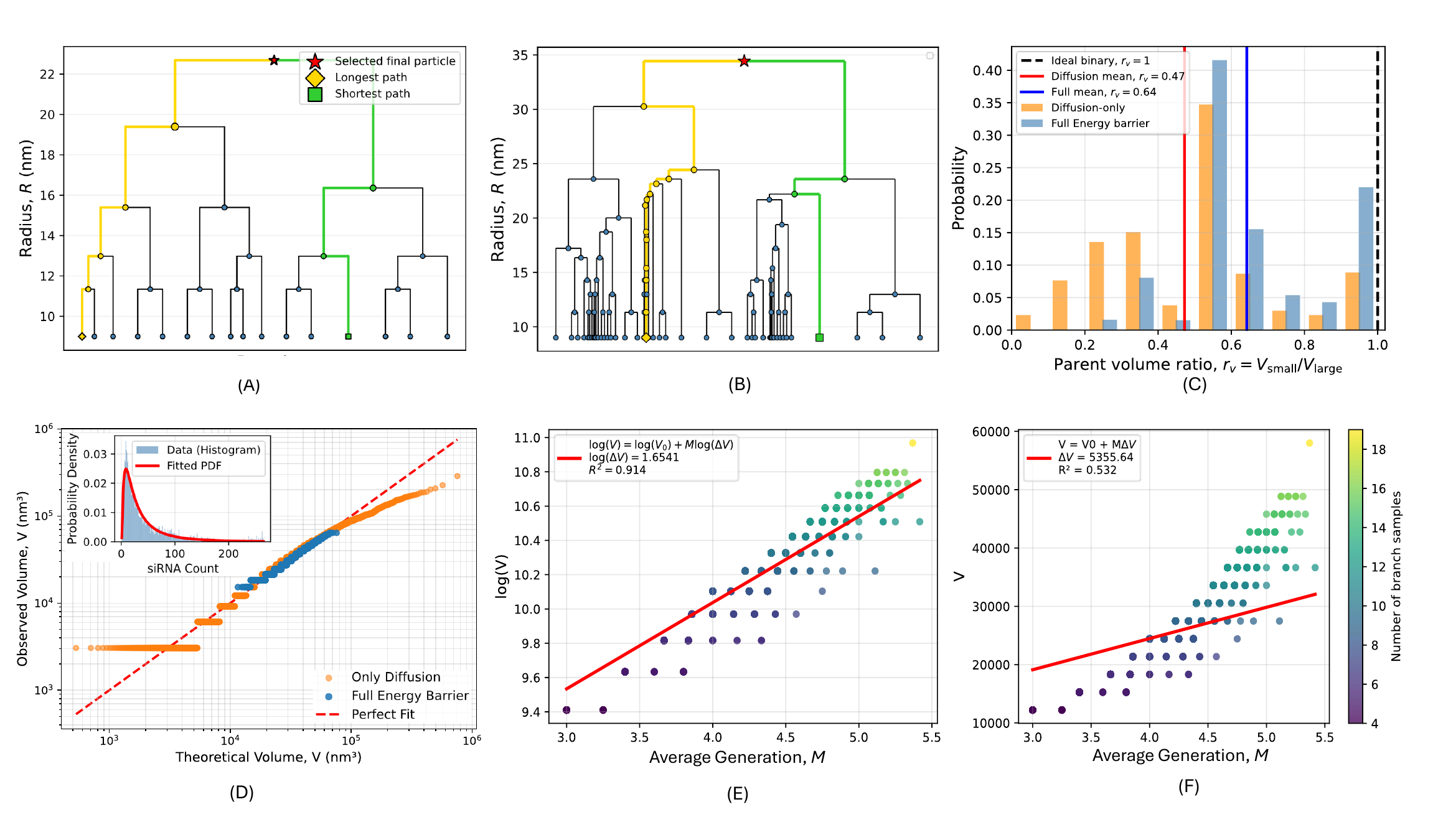}
\caption{\label{fig:lineage_tree}
Lineage-tree analysis of LNP merging pathways from kMC simulations.
(A,B) Representative lineage trees for randomly selected final LNPs using (A) the full energy-barrier model and (B) the diffusion-only model. The full-barrier case appears more binary-like, whereas the diffusion-only case shows more asymmetric side-branch incorporation.
(C) Distribution of the parent volume ratio, $r_v=V_{\rm small}/V_{\rm large}$, for individual merging events. Values near $r_v=1$ indicate equal-size, binary-like merging, whereas values near $r_v=0$ indicate asymmetric large--small merging.
(D) Comparison of simulated LNP volume distributions with log-normal behavior. The full-barrier model gives a narrower distribution, while the diffusion-only model shows broader distributions and stronger deviations at large particle volumes. The inset shows an experimental siRNA payload distribution and log-normal fit from Ref.~\onlinecite{pial_controlling_nodate}.
(E,F) Comparison of multiplicative and additive lineage-scaling models for the full-barrier case. The final LNP volume is better described by multiplicative growth, $V=V_0(\Delta V)^M$, than additive growth, $V=V_0+M\Delta V$, where $\Delta V$ represents an amplification factor in the multiplicative model and a constant volume increment in the additive model.
}
\end{figure*}

For this analysis, we initialized kMC simulations with 4000 LNPs and tracked the complete merging history until only 500 LNPs remained, averaging statistics over 10 independent simulations. Representative lineage trees for randomly selected final LNPs are shown in Fig.~\ref{fig:lineage_tree}(A,B). To assess the role of the interaction barrier, we compared the full energy-barrier model with a diffusion-only model in which all energy barriers were removed ($E_b=0$) and coalescence was controlled only by diffusion-limited encounters.

The full energy-barrier model produces a more binary-like lineage tree. In this case, small LNPs are consumed rapidly because they have lower merging barriers than larger particles, causing them to participate preferentially in early coalescence events. This early depletion of small particles promotes more comparable-size merging at later stages and produces a lineage structure that resembles binary coalescence. In contrast, the diffusion-only model shows stronger asymmetric side-branch incorporation. Without a size-dependent barrier penalty, merging is less selective: small particles can persist to later times and merge directly with much larger particles. Thus, diffusion-only coalescence produces more asymmetric lineages, whereas the full-barrier model favors quasi-binary growth.

To quantify the degree of binary-like merging, we computed the parent volume ratio for each merging event,
\begin{equation}
r_v = \frac{V_{\rm small}}{V_{\rm large}},
\end{equation}
where \(V_{\rm small}\) and \(V_{\rm large}\) are the volumes of the smaller and larger LNPs that merge to form a larger particle. Values near \(r_v=1\) correspond to nearly equal-size, binary-like merging, whereas values near \(r_v=0\) correspond to strongly asymmetric large--small merging. As shown in Fig.~\ref{fig:lineage_tree}(C), the full energy-barrier model shifts the parent-ratio distribution toward larger values, indicating more binary-like merging events. The diffusion-only model has a lower average parent volume ratio, consistent with more asymmetric coalescence pathways.

Figure~\ref{fig:lineage_tree}(D) compares the simulated LNP volume distributions with the log-normal volume distribution. The full-barrier kMC produces a narrower distribution that is closer to log-normal over the range examined. In contrast, the diffusion-only kMC produces a broader distribution and shows stronger deviation from log-normal behavior, particularly for larger particles. This suggests that the energy barrier suppresses excessive asymmetric growth and helps constrain the breadth of the final particle-size distribution. The inset in Fig.~\ref{fig:lineage_tree}(D) shows an experimental siRNA payload distribution from our previous work~\cite{pial_controlling_nodate} which shows a good log-normal fit.

The more binary-like and controlled merging observed in the full energy-barrier model helps explain the emergence of log-normal-like distributions in both LNP volume and payload loading. In a binary-like merging, the particle volume increases by a factor proportional to its current volume at each effective generation, corresponding to a multiplicative growth process. In contrast, in a strongly asymmetric pathway, each step would add approximately the same volume to a growing particle, so the volume increases by an additive growth process. The lineage structure therefore suggests that, under the full energy-barrier model, LNP growth is closer to multiplicative growth than to purely additive accumulation. In an ideal symmetric or binary merging process, each effective merging generation doubles the particle volume, giving
\begin{equation}
V = V_0 2^M,
\end{equation}
where $V_0$ is the initial LNP volume, $V$ is the final LNP volume, and $M$ is the effective number of merging generations. More generally, this relationship can be written as
\begin{equation}
V = V_0(\Delta V)^M,
\end{equation}
where $\Delta V$ is the effective volume amplification factor per merging generation. Taking the logarithm gives
\begin{equation}
\log V = \log V_0 + M\log(\Delta V).
\end{equation}
Thus, if growth is multiplicative, $\log V$ should vary approximately linearly with the effective lineage depth, or generation, $M$.

In contrast, a purely additive growth process would increase the particle volume by a constant volume at each step,
\begin{equation}
V = V_0 + M\Delta V,
\end{equation}
where $\Delta V$ is now the absolute volume increment per merging event. This additive model predicts a linear relationship between $V$ and $M$, rather than between $\log V$ and $M$.

\begin{figure*}[t]
\centering
\includegraphics[width=\textwidth]{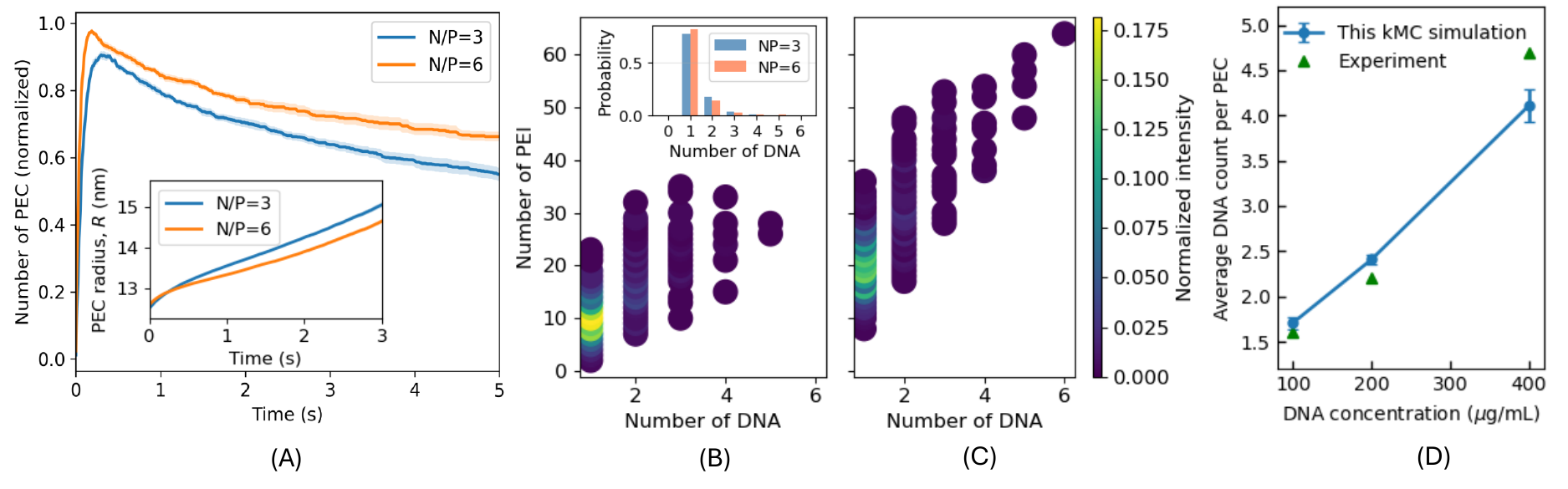}
\caption{\label{fig:pei_dna}
Single-particle kMC predictions for polyelectrolyte complex (PEC) nanoparticle formation by direct complexation of oppositely charged PEI and DNA in water (no solvent exchange). 
(A) Time evolution of the number of PEC particles, normalized with total number of available DNA for two PEI to DNA ratios (N/P=3 and 6). Results show rapid initial complexation followed by slower approach to a steady state.
(B,C) Joint single-particle composition distributions showing the number of PEI chains versus the number of DNA molecules per PEC particle; color indicates normalized probability density. Panels (B) and (C) correspond to N/P=3 and N/P=6, respectively. Inset in (B) shows histogram of DNA count in PEC particles. (D) Average number of DNA molecules per PEC complex as a function of DNA concentration at 5s; error bars denote the standard error of the mean. Results are shown for N/P=3. Experimental data were taken from Hu et al\cite{hu_kinetic_2019-1}.
}
\end{figure*}

Comparison of the kMC lineage statistics with these two models shows that the simulated LNPs with full-energy barrier correlate more strongly with the multiplicative form than with the additive form [Fig.~\ref{fig:lineage_tree}(E,F)]. The fitted amplification factor, $\Delta V=1.65$, is close to the ideal binary-merging limit of $\Delta V=2$, with deviations arising from stochastic pathway variability. In contrast, the additive model shows poor agreement with the lineage statistics, indicating that LNP growth is not well described by constant volume increments. This multiplicative behavior explains the log-normal distributions: if particle volume grows through successive random amplification factors, then $\log V$ becomes a sum of random increments. Consequently, the volume distribution becomes approximately log-normal. Because RNA payload incorporation is coupled to LNP growth and coalescence, the same lineage mechanism also contributes to experimentally observed log-normal-like payload distributions.

\subsubsection{\label{sec:level3}Results for polyelectrolyte complexation: PEI--DNA nanoparticles}


In addition to mRNA--LNP assembly, we demonstrate that the same single-particle kMC framework can be applied to a mechanistically distinct class of multicomponent soft-matter nanoparticles: polyelectrolyte complexes formed by direct electrostatic complexation of oppositely charged macromolecules [Fig.~\ref{fig:epsart}B]. Unlike LNP formation, PEI--DNA complexes form in a common aqueous solvent and do not require solvent-exchange-driven nucleation; instead, nanoparticle growth is initiated upon contact between the polycation and polyanion and proceeds through association, and cluster--cluster aggregation. To reflect pH-dependent charging, we incorporate charge regulation at pH~4, which sets the effective charge of PEI amines and DNA phosphates used in the kMC event rates. Unless otherwise stated, simulations are performed at a DNA concentration of 100~\textmu g/mL using 4.4~kb DNA and 25~kDa PEI, and we vary the nitrogen-to-phosphorus (N/P) ratio, which determines the positive-to-negative charge ratio for fully ionized PEI amines and DNA phosphates, to probe how composition and heterogeneity emerge from the collision history. We assume a turbulent mixing regime.

Figure~\ref{fig:pei_dna} reports particle-resolved PEC kinetics and final composition distributions for two representative states (denoted by N/P=3 and N/P=6). Panel (A) tracks the fraction of PEC complex during complexation, showing a rapid initial transient followed by a slower relaxation as complexes restructure and grow. The number of PEC particles is normalized by the total number of available DNA molecules. Panels (B) and (C) summarize the single-particle stoichiometry by plotting the number of PEI chains versus the number of DNA molecules per complex.  Increasing the N/P ratio broadens the distribution of PEI counts across complexes. Overall, this example illustrates that the framework captures direct complexation systems where pathway dependence arises from charge regulation, diffusion-limited encounters, and stochastic cluster growth.

Figure~\ref{fig:pei_dna}(D) shows the dependence of PEC size on DNA concentration, quantified by the average number of DNA molecules per complex. The mean DNA count per complex increases monotonically as concentration rises from 100 to 400~\textmu g/mL. This trend is expected because higher concentrations increase encounter frequencies and promote cluster--cluster aggregation, shifting the population toward larger complexes. Importantly, the predicted increase in DNA-per-complex with concentration is in good qualitative agreement with existing experimental observations for PEI--DNA polyplex formation \cite{hu_kinetic_2019-1, hou_formation_2011}.

\section{\label{sec:conclusion}Conclusions}

We presented a mechanistic modeling framework for multicomponent nanoparticle assembly that combines early stage assembly with kinetic Monte Carlo modeling of coalescence. The framework takes into account mixing and processing conditions and enables prediction of particle population properties over timescales from microseconds to days. 

Applied to mRNA lipid nanoparticles, the framework shows that processing conditions control the initial conditions entering coalescence and, through collision-driven growth and capture, determine the final heterogeneity in both size and mRNA loading. In particular, the model predicts a strong size–loading correlation that is qualitatively consistent with experimental observations. In addition, lineage analysis of merging events reveals that log-normal-like size and payload distributions emerge from binary-tree-like merging pathways produced by a radius-dependent merging barrier. We further demonstrated the generality of the approach by applying the same single-particle kMC backbone to direct polyelectrolyte complexation of PEI and DNA, a system that does not involve solvent-exchange-driven nucleation. The PEI–DNA example reproduces rapid initial complexation followed by slower aggregate growth and yields particle-resolved stoichiometry distributions that quantify composition heterogeneity beyond ensemble-averaged properties.

Together, these results establish a mechanistic framework for multicomponent nanoparticle assembly that can be adapted across various chemistries by changing interaction rules while retaining the same pathway-resolved single-particle description. This capability provides a route toward rational, process-aware design of nanoparticle formulations based on predicted distributions of size, loading, and composition rather than case-by-case empirical optimization. Open-source code (FormLNP) is provided to facilitate adoption across different systems [https://sites.google.com/view/formlnp/home].

\appendix

\section{Initial growth models at $t < \tau_\tn{m}$}
\label{app:growth_models}

In the main text, the kMC simulations require an initial particle size,
\(R_0\), defined as the average LNP radius at the characteristic solute-mixing time,
\(\tau_m\). This appendix describes two possible models for obtaining \(R_0\): 
(i) a nucleation-and-growth population balance model, and 
(ii) a simplified monodisperse merging-based growth model. Unless otherwise stated, the
results in the main text use the merging-based growth model to estimate the RNA-free LNP
size before the onset of RNA capture and stochastic coalescence.

\subsection{Nucleation and growth population balance model}
\label{app:nucleation_growth}

For solvent-exchange-driven nanoparticle formation, solvent exchange changes the
local molecule solubility and can drive supersaturation, nucleation, and growth \cite{shin_mechanistic_2025}. The solubility of the molecule mixture is estimated using the extended Yalkowsky log-linear model,
\begin{equation}
\ln x_{\text{mix}} =
\varphi_{A_s} \ln x_{A_s}
+
\varphi_{B_s} \ln x_{B_s},
\label{eq:app_solubility}
\end{equation}
where \(x_{\text{mix}}\) is the mole-fraction solubility in the mixed solvent,
\(x_{A_s}\) and \(x_{B_s}\) are the solubilities in pure solvents \(A_s\) and \(B_s\),
and \(\varphi_{A_s}\) and \(\varphi_{B_s}\) are the corresponding solvent volume fractions.

The temporal evolution of the particle number density distribution \(n(L,t)\), where
\(L\) is the particle diameter, is described by a population balance equation \cite{iggland_population_2012, ramkrishna2000population, vetter_modeling_2013},
\begin{equation}
\frac{\partial n}{\partial t}
=
B_n
-
\frac{\partial \left[G(L,t)n\right]}{\partial L}.
\label{eq:app_pbe}
\end{equation}
Here, \(B_n\) is the nucleation birth rate and \(G(L,t)\) is the size-dependent growth
rate. Although the population balance is formulated in terms of diameter, results are
reported in terms of particle radius, \(R=L/2\).

The critical nucleus size can be obtained from classical nucleation theory as
\begin{equation}
L_c =
\frac{4 \gamma V_m}{k_B T \ln S},
\label{eq:app_critical_size}
\end{equation}
where \(\gamma\) is the interfacial tension, \(V_m\) is the molecular volume of the lipid,
\(k_B\) is Boltzmann's constant, \(T\) is the absolute temperature, and
\(S=C/C_{\mathrm{eq}}\) is the supersaturation ratio. The corresponding nucleation rate is
\begin{equation}
J =
A_n
\exp\left(
-\frac{\Delta G_c}{k_B T}
\right),
\label{eq:app_nucleation_rate}
\end{equation}
with the critical nucleation barrier
\begin{equation}
\Delta G_c =
\frac{16 \pi \gamma^3 V_m^2}
{3\left(k_B T \ln S\right)^2}.
\label{eq:app_critical_barrier}
\end{equation}
Here \(A_n\) is a kinetic prefactor.

To avoid placing all newly nucleated particles at a single diameter, the birth rate can be
distributed around the critical diameter using a normalized Gaussian kernel,
\begin{equation}
B_n(L,t)
=
J(t)\mathcal{G}(L,L_c),
\label{eq:app_birth_rate}
\end{equation}
where
\begin{equation}
\mathcal{G}(L,L_c)
=
\frac{1}{\mathcal{N}}
\exp\left[
-\frac{(L-L_c)^2}{2\sigma_n^2}
\right],
\label{eq:app_gaussian_kernel}
\end{equation}
and \(\mathcal{N}\) is chosen such that
\(\int \mathcal{G}(L,L_c)\,dL=1\). Therefore,
\(\int B_n(L,t)\,dL=J(t)\).

Particle growth is driven by the difference between the bulk solute concentration and the
size-dependent equilibrium concentration. The Gibbs--Thomson relation gives
\begin{equation}
S^*(L)
=
\frac{C^*(L)}{C_{\mathrm{eq}}}
=
\exp\left(
\frac{4\gamma V_m}{k_B T L}
\right).
\label{eq:app_size_dep_sat}
\end{equation}
The growth rate is written as
\begin{equation}
G(L,t)
=
k_g C_{\mathrm{eq}} L^{\beta}
\left[
S(t)-S^*(L)
\right]^{\alpha},
\label{eq:app_growth_rate}
\end{equation}
where \(k_g\) is the growth-rate constant, and \(\alpha\) and \(\beta\) determine the
growth mechanism.

Equation~\eqref{eq:app_pbe} can be discretized on a finite-volume grid in particle diameter.
The growth term then can be evaluated with an upwind scheme to ensure numerical stability and to
avoid spurious oscillations during advective transport in size space. The resulting system
of ordinary differential equations can then be integrated using adaptive time stepping. The particle
size at the mixing time, \(R(\tau_m)\), can then be used as the initial radius for the kMC
coalescence simulations.

\subsection{Monodisperse merging-based LNP growth model}
\label{app:merging_growth}

As an alternative to the population balance model, we also use a simplified monodisperse
growth model to describe LNP growth before RNA capture. In this model, LNPs grow during
the interval \(t<\tau_m\) through RNA-free LNP--LNP merging. The particle population is treated as monodisperse using a mean-field approximation, so the model tracks only the average particle radius
\(R(t)\). The radius at the solute-mixing time,
\begin{equation}
R_0 = R(\tau_m),
\end{equation}
is then used as the initial particle size in the kMC simulations.

The mean field approximation of merging rate per particle is assumed to depend on the particle
concentration, the diffusion-limited collision kernel, and an Arrhenius
factor that accounts for the interaction barrier:
\begin{equation}
k_m(R)
=
K_{\mathrm{coll}}(R)c(R)
\exp\left[-\frac{E_b(R)}{k_BT}\right].
\label{eq:km_general}
\end{equation}
Here \(c(R)\) is the number concentration of LNPs of radius \(R\),
\(K_{\mathrm{coll}}(R)\) is the diffusion-limited collision kernel, and
\(E_b(R)\) is the fusion energy barrier for RNA-free LNPs. The barrier is
taken as the sum of steric PEG--PEG repulsion and DLVO interactions,
\begin{equation}
E_b(R)=E_{\mathrm{PEG}}(R)+W_{\mathrm{DLVO}}(R),
\label{eq:E_total}
\end{equation}
where \(W_{\mathrm{DLVO}}\) is evaluated using the DLVO interaction model
described in the main text, but without RNA contributions to the
charge-regulation calculation.

For a monodisperse population, the particle number concentration is
estimated from conservation of lipid volume. If \(c_0\) is the lipid
number concentration and \(v_0\) is the volume per lipid molecule, then
\begin{equation}
c(R)V_1(R)=c_0v_0,
\end{equation}
or
\begin{equation}
c(R)=\frac{c_0v_0}{V_1(R)}.
\label{eq:cR}
\end{equation}
Here
\begin{equation}
V_1(R)=\frac{4\pi R^3}{3}
\label{eq:V1}
\end{equation}
is the volume of one LNP. Therefore,
\begin{equation}
c(R)
=
\frac{3c_0v_0}{4\pi R^3}.
\label{eq:cR_explicit}
\end{equation}

The Stokes--Einstein diffusivity of a particle of radius \(R\) is
\begin{equation}
D(R)=\frac{k_BT}{6\pi\eta R},
\label{eq:stokes_einstein}
\end{equation}
where \(\eta\) is the solvent viscosity. For diffusion-limited collisions
between two spherical particles, the Smoluchowski collision kernel is
\begin{equation}
K_{\mathrm{coll}}
=
4\pi(D_1+D_2)(R_1+R_2).
\end{equation}
For two identical LNPs, \(D_1=D_2=D(R)\) and \(R_1=R_2=R\), so
\begin{equation}
K_{\mathrm{coll}}(R)
=
4\pi[2D(R)][2R]
=
16\pi D(R)R.
\label{eq:Kcoll}
\end{equation}
Substituting Eqs.~\eqref{eq:cR_explicit} and \eqref{eq:Kcoll} into
Eq.~\eqref{eq:km_general} gives
\begin{align}
k_m(R)
&=
16\pi D(R)R
\frac{3c_0v_0}{4\pi R^3}
\exp\left[-\frac{E_b(R)}{k_BT}\right] \\
&=
\frac{12D(R)c_0v_0}{R^2}
\exp\left[-\frac{E_b(R)}{k_BT}\right].
\end{align}
Using Eq.~\eqref{eq:stokes_einstein}, this becomes
\begin{equation}
k_m(R)
=
\frac{2k_BT c_0v_0}{\pi\eta R^3}
\exp\left[-\frac{E_b(R)}{k_BT}\right].
\label{eq:km_final}
\end{equation}
Thus, Eq.~\eqref{eq:km_final} describes diffusion-controlled encounters
between equal-sized LNPs multiplied by the Arrhenius probability of
successful fusion.

Each successful merging event combines two particles into one larger
particle. Because two particles are consumed per merger, the average
particle volume evolves according to
\begin{equation}
\frac{\partial V_1}{\partial t}
=
\frac{k_m(R)V_1}{2}.
\label{eq:dVdt}
\end{equation}
The factor of \(1/2\) accounts for the stoichiometry of binary merging.
Using \(V_1=4\pi R^3/3\),
\begin{equation}
\frac{\partial V_1}{\partial t}
=
4\pi R^2\frac{\partial R}{\partial t}.
\end{equation}
Substituting this into Eq.~\eqref{eq:dVdt} gives
\begin{equation}
4\pi R^2\frac{\partial R}{\partial t}
=
\frac{k_m(R)}{2}
\frac{4\pi R^3}{3}.
\end{equation}
Therefore,
\begin{equation}
\frac{\partial R}{\partial t}
=
\frac{k_m(R)R}{6}.
\label{eq:dRdt_km}
\end{equation}
Finally, substituting Eq.~\eqref{eq:km_final} into
Eq.~\eqref{eq:dRdt_km} yields
\begin{equation}
\frac{\partial R}{\partial t}
=
\frac{k_BT c_0v_0}{3\pi\eta R^2}
\exp\left[-\frac{E_b(R)}{k_BT}\right].
\label{eq:dRdt_final}
\end{equation}

This is integrated from the initial particle
radius \(R(t_0)=3 \)nm (approximately the radius of small lipid micelles) to the mixing time \(\tau_m\). The resulting radius
defines the initial LNP radius used in the kMC simulations. This model
neglects polydispersity before \(\tau_m\), but provides a compact way to
incorporate the effects of lipid concentration, PEG-lipid steric
stabilization, electrostatic interactions, and flow-dependent mixing time
into the kMC initial condition.

\begin{acknowledgments}
This work was supported by start-up funds provided by the Whiting School of Engineering at JHU to TC. Computational work was carried out at the Advanced Research Computing at Hopkins (ARCH) core facility (rockfish.jhu.edu), which is supported by the National Science Foundation (NSF) grant number OAC 1920103.
\end{acknowledgments}

\section*{Data Availability Statement}
The data that support the findings of this study are available from the corresponding author upon reasonable request.
\nocite{*}
\bibliography{jcp}

\end{document}